\documentclass[aps,prb,twocolumn,groupedaddress,floatfix,showpacs]{revtex4}

\usepackage{graphics}
\usepackage{epsfig}
\usepackage[latin1]{inputenc}
\usepackage{subfigure}
\usepackage{dcolumn}
\usepackage{bm}
\usepackage{overpic} 

\begin{document}

\title{Characterization of amorphous and crystalline silicon nanoclusters in ultra thin silica layers.}
\author{Annett Thogersen}
\author{Jeyanthinath Mayandi}
\author{Terje G. Finstad}

\affiliation{Centre for Materials Science and
  Nanotechnology, University of Oslo, P.O.Box 1126 Blindern, N-0318 Oslo, Norway}

\author{Arne Olsen}
\author{Jens Sherman Christensen}
\affiliation{Department of Physics, University of Oslo, P.O.Box 1048 Blindern, N-0316 Oslo, Norway}

\author{Masanori Mitome}
\author{Yoshio Bando}

\affiliation{National Institute of Material Science, Namiki 1-1, Tsukuba, Ibaraki, 305-0044 Japan}

\date{\today}

\begin{abstract}

\noindent The nucleation and structure of silicon nanocrystals formed by different preparation conditions and silicon concentration (28 - 70 area \%) have been studied using Transmission Electron Microscopy (TEM), Energy Filtered TEM (EFTEM) and Secondary Ion Mass Spectroscopy (SIMS).  The nanocrystals were formed after heat treatment at high temperature of a sputtered 10 nm thick silicon rich oxide on 3 nm SiO$_2$ layer made by Rapid Thermal Oxidation (RTO) of silicon. Nanocrystals precipitate when the excess silicon concentration exceeds 50 area \%. Below this percentage amorphous silicon nanoclusters were found. In-situ heat treatment of the samples in the TEM showed that the crystallization requires a temperature above 800$^o$C. The nanocrystals precipitate in a 4 nm band, 5 nm from the Si substrate and 4 nm from the SiO$_2$ sample surface. The silicon nucleates where the excess Si concentration is the highest. The top surface has less excess Si due to reaction with oxygen from the ambient during annealing. The SiO$_2$-RTO layer is more Si rich due to Si diffusion from the SiO$_2$-Si layer into RTO. Twinning and stacking faults were found in nanocrystals with 4-10 nm in diameter. These types of defects may have large effects upon the usability of the material in electronic devices. Both single and double twin boundaries have been found in the nanocrystals by high resolution transmission electron microscopy (HRTEM). Image simulations were carried out in order to obtain more information about the defects and nanocrystals. The stacking faults are extrinsic and located in the twin boundaries.

\end{abstract}



\maketitle

\section{Introduction}

\noindent Nanoscaled electronic devices have attracted much attention, especially related to MOS (Metal-Oxide-Semiconductor) devices used for memory storage applications \cite{beltsios:mos, boer:mos}. The improved electrical properties and the possibility to scale electronic devices down considerably are important and useful. Silicon nanocrystals (NCs) embedded in silica can potentially be used for various applications \cite{pavesi:si} such as nanocrystal memory cells \cite{tiwari:si}, photon converters, optical amplifiers etc \cite{han:intro, kik:intro}. It is assumed that the replacement of the bulk floating gate with nanocrystals will in turn result in longer retention, lower gate voltage and lower power consumption \cite{chen:shift}. The separation between the nanocrystals can prevent charge loss laterally and results in short writing times at lower voltages and improved reliability \cite{ferdi:bias}. Accurate control of the array of nanocrystals is very important, since even 1 nm change in tunnel distance can affect the write and erase time \cite{tof:perego}. Both injection and retention of electrons in these devices are very sensitive to the size, distribution, interfaces and electronic structure of the nanocrystals. It is desirable to make nanocrystals that are less than 10 nm in diameter so that the Coulomb blockade effect becomes prominent at room temperature \cite{asi:1}. When the dimensions approach the atomic scale, significant changes occur in the electronic, optical and thermodynamic properties compared to bulk materials \cite{ion:1}. Therefore it is important to get information about the nucleation of amorphous and crystalline nanoclusters, to characterize their atomic and electronic structure, the nanocrystal interfaces, and the defects within and around the nanocrystals.

Twins are among the most common defects in Si \cite{wang:twin} and depend on the stacking fault energy of the material, the surface stresses and surface orientations \cite{park:2}. Low stacking fault energy results in an increase in twinning \cite{park:2}. Nanocrystal twinning can be due to stresses in and on the surface of nanocrystals, coalescence of smaller nanocrystals or it can occur during growth and heat treatment. Wang et al.\cite{wang:twin} studied nanocrystals made by ion implantation of Si in a 1$\mu$m SiO$_2$ film. They found twinning and stacking faults in 90 \% of the nanocrystals with a crystal size larger than 5 nm. They claimed that nanocrystals smaller than 5 nm will not contain any twinning or stacking faults \cite{wang:stacking, wang:twin}. Perrey et al.\cite{perry:defect} studied silicon nanocrystal defects, precipitated from hydrogenated amorphous thin Si films (a/nc-Si:H). Both twinning and stacking faults were found in the silicon nanocrystals with sizes 1.5 - 5 nm in diameter. Few similar studies have been done on nanocrystals made by sputtering in ultra-thin oxides (less than 20 nm in thickness).

Most previous studies of 3-12 nm  nanocrystals have been carried out on thick silicon oxides (500nm - 1$\mu$m) with implantation doses between 2*10$^{16}$ - 3*10$^{17}$ Si$^+$ cm$^{-2}$ (3-50 at. \%) \cite{wang:1, ion:1, nicklaw:defect, wang:stacking, wang:twin, ross:si, mayandi:nano}. Si nanocrystal formation, size and distribution in thin oxides have previously not been studied in detail. The nucleation of Si nanocrystals depends on the density of the SiO$_2$ layer and defects. This varies with different sample preparation techniques. Nanocrystals have previously been made by methods like sputtering, sol-gel, ion implantation and other chemical processes \cite{ion:1}. The nanocrystals in the present work were intended to be formed in a 10-15 nm thin oxide and close to the surface. The concentration and nanocrystal size in ion implanted samples have previously been shown to have a Gaussian distribution\cite{wang:stacking, ross:si} with a peak position further down in the oxide. Ion implantation could not be used on such thin oxides, due to less control of the average cluster size and density \cite{ion:1}. Therefore sputtering was used. The nucleation of nanocrystals, distribution, crystal size and the defects were studied in the present work by High Resolution Transmission Electron Microscopy (HRTEM), Energy Filtered TEM (EFTEM) and Secondary Ion Mass Spectroscopy (SIMS). In order to get detailed information about the nanocrystals and their defects, experimental through focus series were compared to simulated images.

\section{Experimental}

\noindent The samples were made by growing a ~3 nm thin layer of SiO$_2$ on a p-type silicon substrate by Rapid Thermal Oxidation (RTO) at $1000^oC$ for 6 sec. Then a ~10 nm layer of silicon rich oxide was sputtered from a SiO$_2$:Si composite target onto the RTO and heat-treated in a N$_2$ atmosphere at $1000-1100^oC$ for 30-60 min. Different area percentages (area coverage) of Si:SiO$_2$ (6, 8, 17, 28, 42,  50, 60, 70 area \% corresponding to 4, 5, 11, 18, 33, 40, 46 at. \%) were used to produce different silicon supersaturations in the oxide. The samples dicussed in this paper are given in Table \ref{table:exp}. Cross-sectional TEM samples were prepared by ion-milling using a Gatan precision ion-polishing system operated at 5 kV gun voltage. Silicon nanocrystals were observed by HRTEM and Energy Filtered imaging of the plasmon peak. The HRTEM and Energy Filtered TEM (EFTEM) were performed with a 300 keV JEOL 3100FEF microscope with an Omega energy filter. The spherical (C$_s$) and chromatic aberration (C$_c$) coefficients of the objective lens were 0.6 mm and 1.1 mm, respectively. The point to point resolution was 0.174 nm at Scherzer focus (-37 nm), and the minimum probe diameter was 0.2 nm. The energy resolution at 300 keV was 0.78 eV, and was previously determined experimentally \cite{mitome:exp}. The energy dispersion of the Omega-filter was 0.85 $\mu$m/eV at 300 keV. Through focus series of HRTEM images of crystal orientation $[1\bar{1}0]$ were recorded with an objective aperture large enough to include the 220 reflection (0.192nm) corresponding to 5.6 nm$^{-1}$ (11 mrad) radius. Ten images with a 10 nm difference in focus were obtained around Scherzer condition. The images were compared to simulated focus series made by using the MacTempas computer program. In-situ heating of sample 70asd (see Table \ref{table:exp}) was in addition carried out with the JEOL 3100FEF microscope using a heating holder. EFTEM- Spectral Imaging (EFTEM-SI) was acquired by scanning from 2 eV to 30 eV energy losses, with an energy slit of 2 eV. Electron energy loss spectra were aquired with a 197 keV JEOL 2010F microscope equipped with a Gatan imaging filter and detector. Secondary ion mass spectrometry (SIMS), using a Cameca IMS 7f with a 25 nA O$_2^{+}$, 0.5 keV primary beam was used to obtain depth profiles of Si secondary intensities. The ion beam was scanned over an area of 200 x 200 $\mu$m$^2$. Positive secondary ions were collected from a circular area with a diameter of 62 $\mu$m from the center of the crater and counted with an electron multiplier. After the SIMS measurements the depth of the craters were measured with a Dektak 8 stylus profilometer, for converting the sputter time to depth. The depth calibration was made on the assumption of a constant erosion rate.

\begin{table}[hp]
\caption{The samples discussed in this paper, presented with Si concentration, heating time and temperature.}
\begin{tabular}{cccc}
\hline
\hline
sample & area \% & heating & heating \\
name & Si & time & temp \\
 & & (min) & ($^o$C) \\
\hline
28 & 28 & 1000 & 30 \\
42 & 42 & 1000 & 60 \\
50 & 50 & 1000 & 30 \\
70 & 70 & 1100 & 60 \\
70asd\footnotemark[1] & 70 & 0 & 0 \\
\hline 
\hline
\end{tabular}
\footnotetext[1]{as deposited}
\label{table:exp}
\end{table}

\section{Results and discussion}

\subsection{Nucleation of crystalline and amorphous nanoclusters}

\begin{figure}
  \begin{center}
    \includegraphics[width=0.4\textwidth]{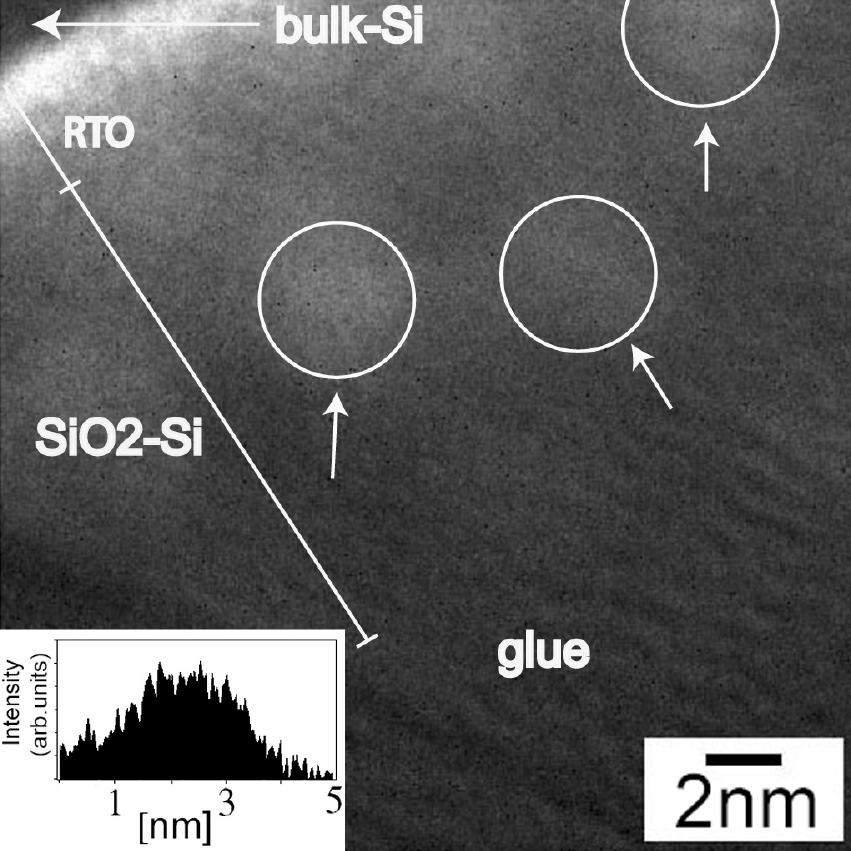}
    \caption{EFTEM image (at 16.8 eV) of amorphous nanoclusters in sample 42. The arrows indicates the amorphous nanoclusers which are Si rich. The graph in the left corner shows the differences in intensity over the bright dot in the middle.}
    \label{figure:1}
  \end{center}
\end{figure}

\noindent The nucleation of nanocrystals with different preparation conditions was studied by HRTEM and EFTEM. No silicon nanocrystals were found in sample 28 (see Table \ref{table:exp}). At silicon fraction of 42 \% (sample 42) 4-6 nm bright, isolated areas with abrupt interfaces are visible in the EFTEM image shown in Figure \ref{figure:1}. The graph in the left corner shows how the contrast changes across one of the bright area in the middle of the image. The bright areas are located about 5 nm from the silicon substrate and 4 nm from the specimen surface (SiO$_2$/glue). The EFTEM image was produced by using the plasmon Si peak at 16.8 eV. HRTEM images show no visible nanoclusters or any contrast differences in this area. The bright, isolated areas in Figure \ref{figure:1} of sample 42 are probably amorphous Si nanoclusters, because the local changes in image intensity cannot be explained by variations in sample thickness. Since no nanoclusters were found in sample 28, the excess silicon had either oxidized, precipitated to small amorphous nanoclusters or are still silicon atoms or ions in solid solution. EFTEM images of the surface plasmon peak of Si were also acquired, but no such bright areas were visible.

\begin{figure}
  \begin{center}
    \includegraphics[width=0.4\textwidth]{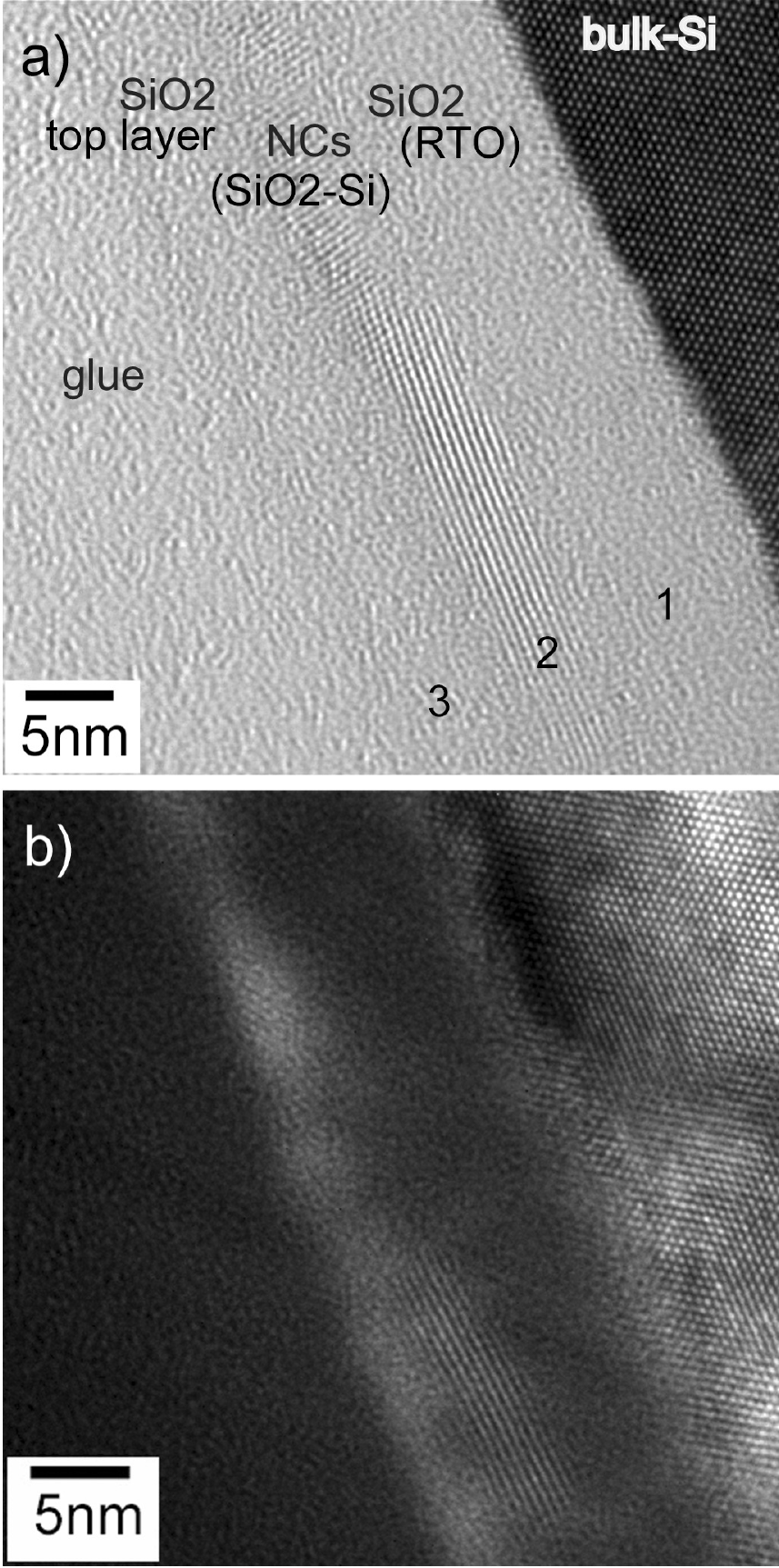}
    \caption{Images of sample 70. a) HRTEM image, b) EFTEM image, with the plasmon peak of silicon (16.8 eV). The numbers indicate the area where the EDS and EELS spectra were obtained (see Figure \ref{figure:4}).}
    \label{figure:2}
  \end{center}
\end{figure}

HRTEM and EFTEM images of samples 50 and 70 show areas with lattice fringes indicating crystalline nanoclusters (see Figure \ref{figure:2}). The nanocrystals in these samples precipitate at the same distance from the substrate as the amorphous nanoclusters in sample 42. The nanocrystals first grow spherically up to about 4 nm in diameter, following an elongated growth laterally in the $<$111$>$ directions (see Figure \ref{figure:2}). The nanocrystals have the same diamond type structure as bulk silicon with lattice parameter 0.54 nm. These observations show that increasing silicon concentration, annealing time or temperature induce nanocrystal growth.

\begin{figure}
  \begin{center}
    \includegraphics[width=0.4\textwidth]{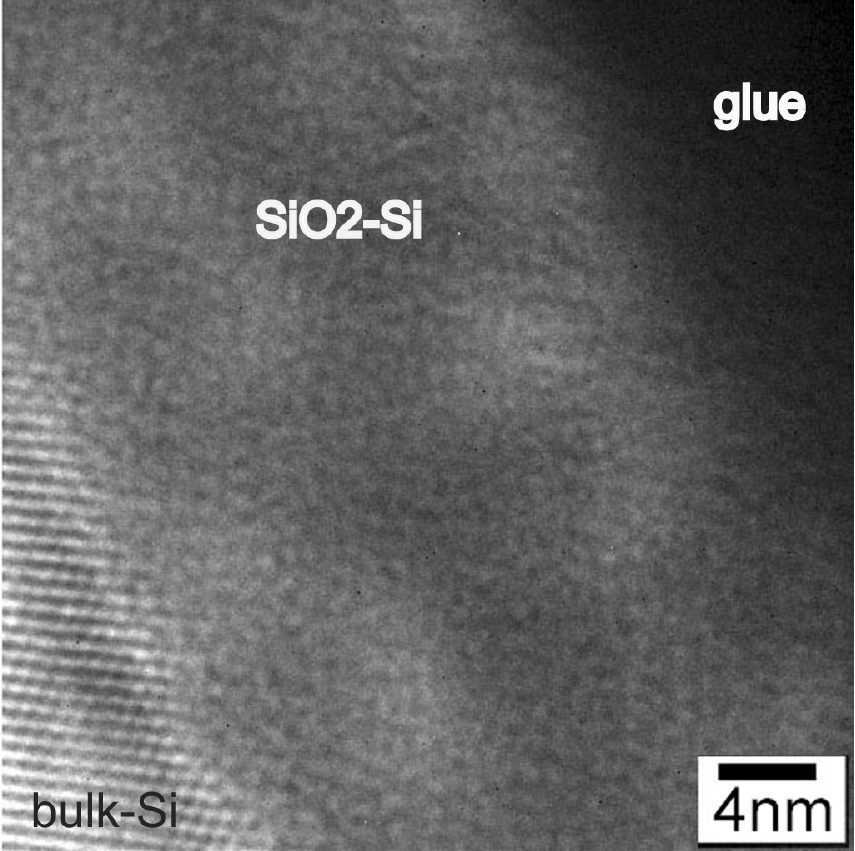}
    \caption{EFTEM-SI of sample 70asd after an in-situ heat treatment for 1 hour at 800$^o$C. The image is made with the plasmon peak of Si (16.8 eV).}
    \label{figure:3}
  \end{center}
\end{figure}

Sample 70asd was examined with HRTEM and EFTEM-SI. No lattice fringes were observed in the HRTEM images. With EFTEM-SI small bright areas were observed. After an in-situ heat treatment in the TEM for 1 hour at $800^o$C, still no indications of crystalline nanoclusters were found (see Figure \ref{figure:3}). The bright areas in the EFTEM images are probably small amorphous Si nanoclusters. During heat treatment, the Si atoms in the nanoclusters diffuse and form larger areas of amorphous Si. Crystalline nanoclusters were only formed after heating at a temperature above $800^o$C, see Figure \ref{figure:2}.

\begin{figure}
  \begin{center}
    \includegraphics[width=0.4\textwidth]{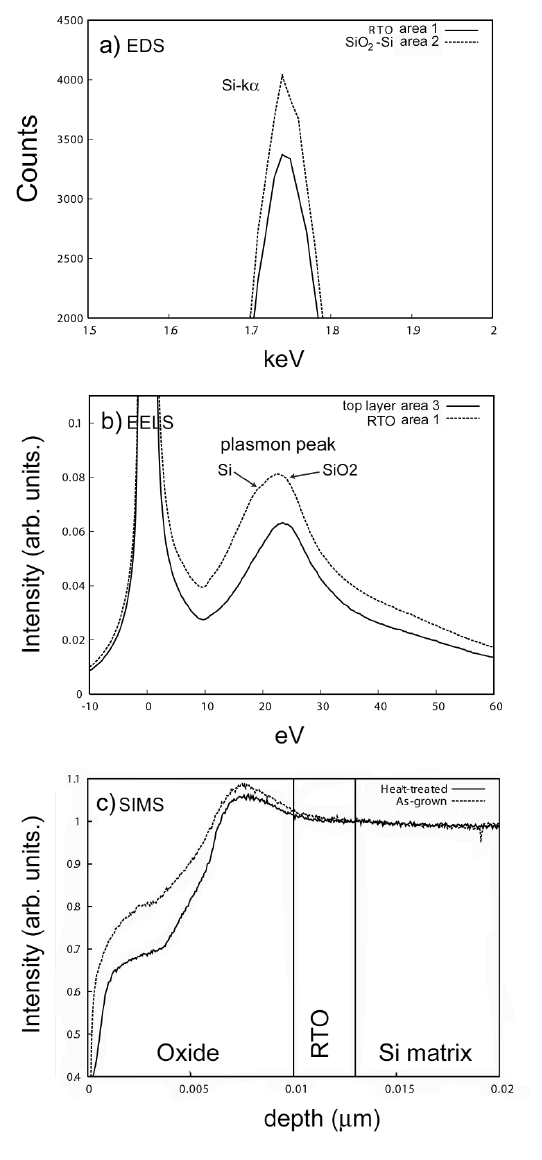}
    \caption{EDS, EELS and SIMS measurements on sample 70 and 70asd. The location of the measurements are shown in Figure \ref{figure:2}a. a) EDS spectra of the silicon K$\alpha$ line are from the SiO$_2$-RTO (area 1) and the SiO$_2$-Si (area 2). b) EELS spectra of the SiO$_2$-RTO (area 1) and the top layer (area 3). The two spectra are normalizes with regard to the zero loss peak. c) SIMS data from sample 70 and 70 asd. The spectra are normalized so that the Si matrix has a secondary ion intensity of 1.0.}
    \label{figure:4}
  \end{center}
\end{figure}

EDS spectra of the SiO$_2$-RTO and the SiO$_2$-Si layer in sample 70 are presented in Figure \ref{figure:4}a. The location of the EDS measurements are shown in Figure \ref{figure:2}a. Both EDS spectra were acquired with the same spot size and acquisition time. The Si-K$\alpha$ EDS peak have almost the same intensity for both regions, but slightly higher from the SiO$_2$-Si area than the RTO. The latter area (RTO) has slightly higher sample thickness compared to the SiO$_2$-Si (area 2) layer. Therefore the differences in EDS intensities indicate that the RTO layer is probably not only composed by SiO$_2$, but contain high amounts of pure Si. 

The increase in silicon concentration in the RTO compared to pure SiO$_2$ can also be seen from the EELS spectra in Figure \ref{figure:4}b. These spectra were acquired from the SiO$_2$-RTO (area 1) and the top layer (area 3), see Figure \ref{figure:2}a. The RTO is slightly thicker than the top layer, and therefore has a higher plasmon peak (as seen in Figure \ref{figure:4}b). There is a small difference (1.2 eV) in plasmon peak energy between the RTO (22.8 eV) and the top layer (24.0 eV). The difference is within the standard deviation of the spectrometer (1.5 eV), but the experimental observed energy shift may indicate trends. The plasmon peak energy of pure silicon is 16.8 eV, while pure SiO$_2$ has 24 eV. A shift towards lower plasmon peak energy may therefore indicate an increase in silicon concentration in the oxide. The plasmon peak of the RTO exhibits in addition a broader width in the range of 19-25 eV. This is due to a significant increase in the pure silicon contribution (at 16.8 eV) in the oxide \cite{nicotra:eels}. 

SIMS data in Figure \ref{figure:4}c shows the Si secondary intensity profile of sample 70 and 70asd. The Si concentration maximum is located 7 nm from the sample surface. This could be due to oxygen diffusion from the surface into the oxide. There is also an increased Si concentration in the SiO$_2$-RTO (area 1) than may be expected from pure SiO$_2$. This could be due to silicon diffusion from the SiO$_2$-Si layer (area 2) into the SiO$_2$-RTO layer. Sample 70 and 70asd have different Si secondary ion profiles. This suggests that diffusion has occurred both before and after heat treatment. The observed Si secondary intensity is higher in the oxide than in the bulk due to a matrix effect \cite{lasse:sims}. The amorphous and crystalline nanoclusters both nucleate at the concentration maximum. This maximum is located about 7 nm below the surface and is due to both oxygen and silicon diffusion before and after the heat treatment. When sample 70 and 70asd is exposed to air, oxygen diffuses into the top layer (area 3) and additional silicon diffuses from the SiO$_2$-Si or from the Si-substrate into the RTO. This is in agreement with both the EDS, EELS and SIMS data in Figure \ref{figure:4}.

\subsection{Analysis of HRTEM images}

\begin{figure}
  \begin{center}
    \includegraphics[width=0.4\textwidth]{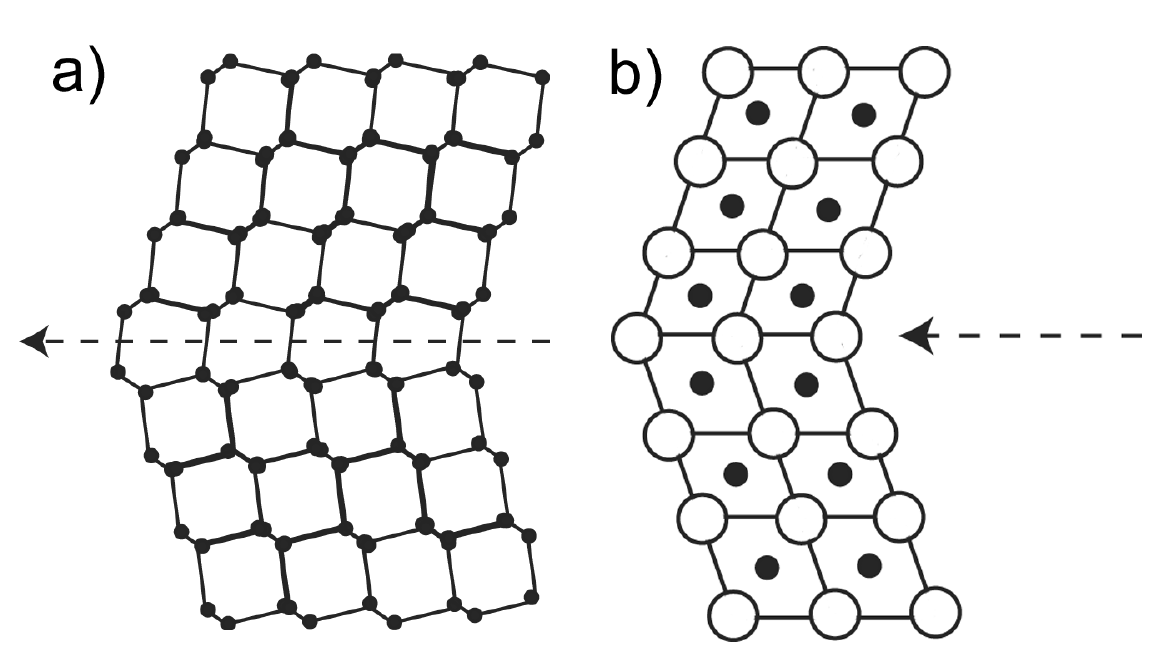}
    \caption{ a) A (111) twin boundary in a diamond type structure, projected along [01$\overline{1}$] and b) a simplified projection seen at limited resolution.}
    \label{figure:5}
  \end{center}
\end{figure}

\begin{figure}
  \begin{center}
    \includegraphics[width=0.4\textwidth]{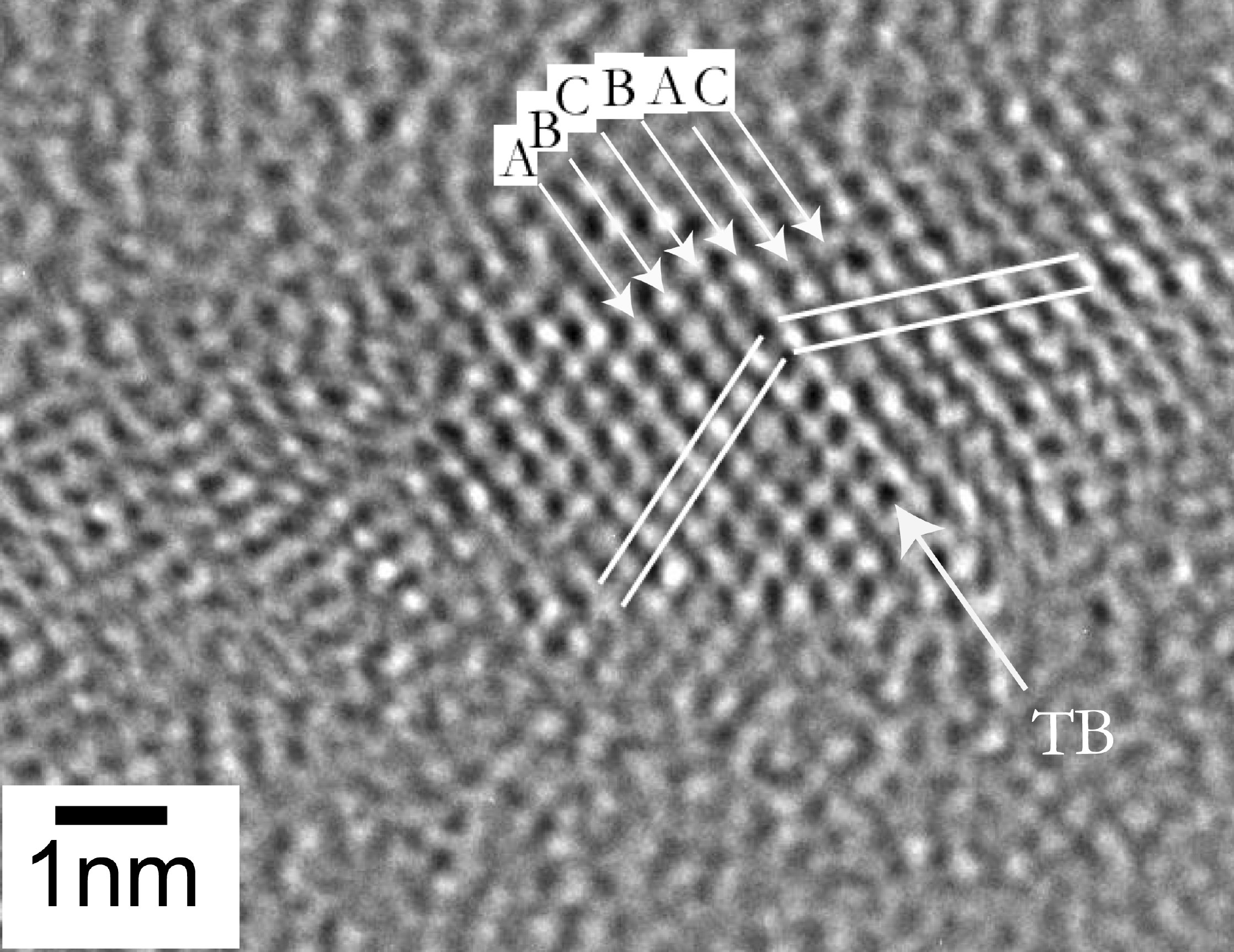}
    \caption{Single-twin in a nanocrystal in sample 70.}
    \label{figure:6}
  \end{center}
\end{figure}

\begin{figure}
  \begin{center}
    \includegraphics[width=0.4\textwidth]{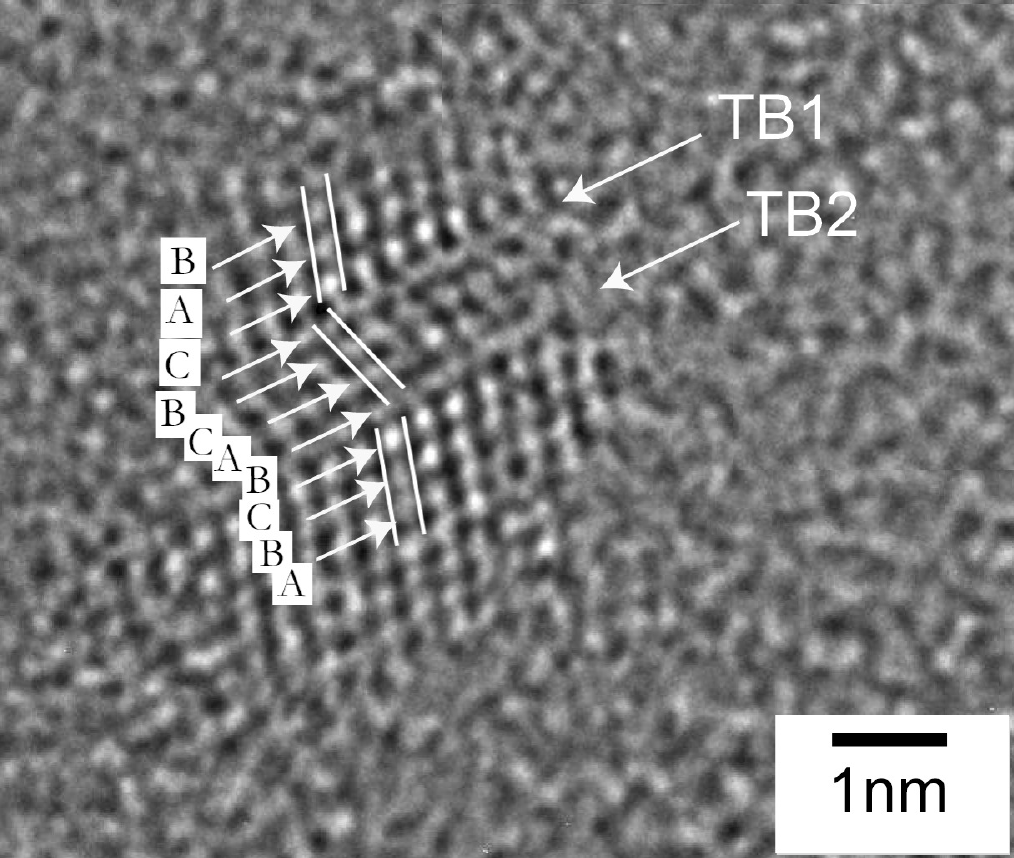}
    \caption{Double-twin in a nanocrystal in sample 70.}
    \label{figure:7}
  \end{center}
\end{figure}

\begin{figure}
  \begin{center}
    \includegraphics[width=0.4\textwidth]{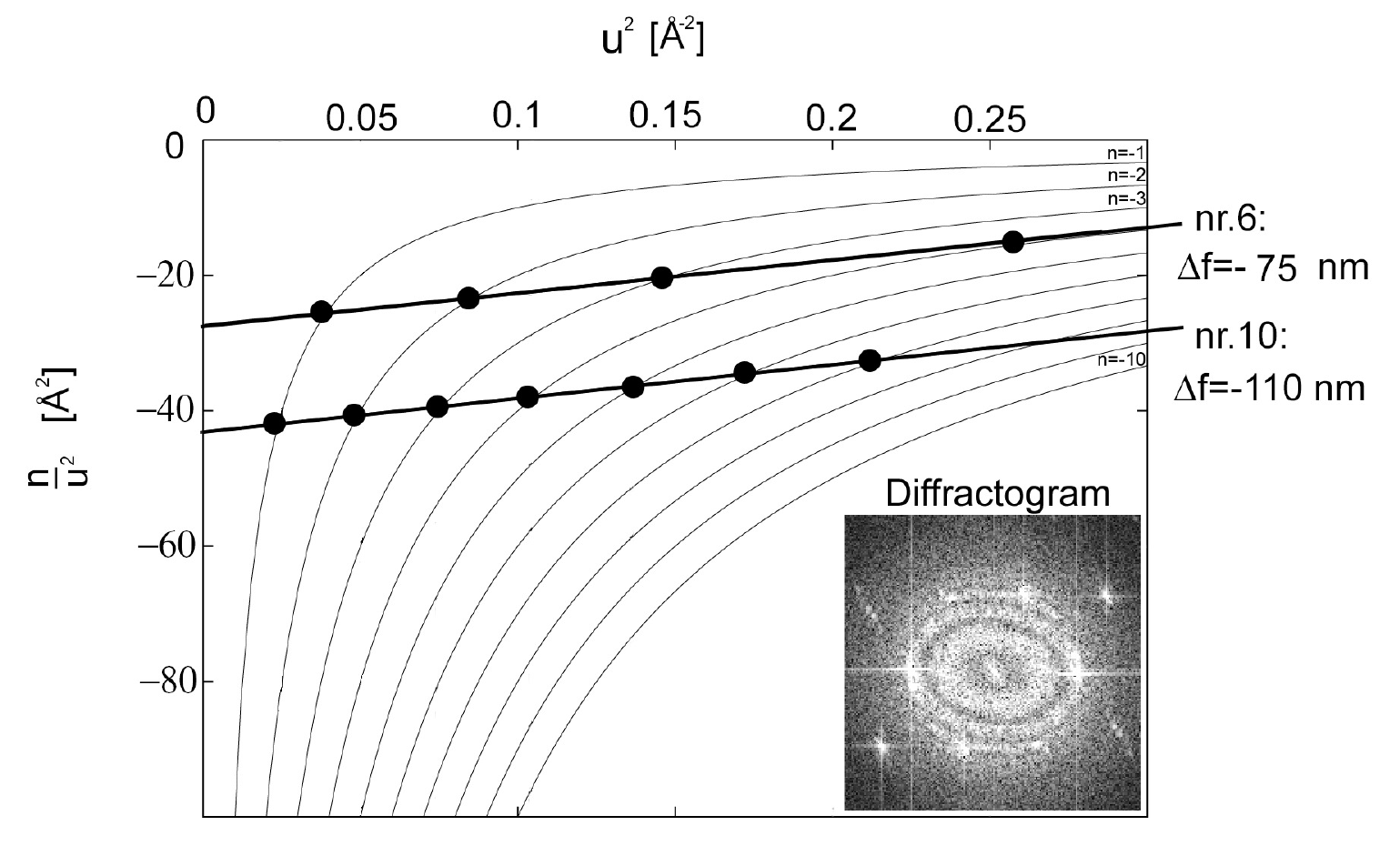}
    \caption{A plot of $n/{u^2}$ as a function of u$^2$. The plot is used to determine the focus of the HRTEM images in Figure \ref{figure:9}. The diffractogram in the bottom right corner corresponds to HRTEM image no. 10 (see Figure \ref{figure:9}). The spots in the diffractogram from the Si substrate was used as internal standard.}
    \label{figure:8}
  \end{center}
\end{figure}

\noindent In HRTEM images of thin silicon specimens, bright dots at Scherzer conditions correspond to the tunnels in the structure\cite{Spence:gammel}. In thicker regions the bright dots in the images correspond to the atomic columns. To distinguish between these two cases in a perfect crystal, is impossible based only on the image contrast \cite{Spence:gammel}. Figure \ref{figure:5}a shows a schematic drawing of a (111) twin boundary in a diamond type structure in the [0$\overline{1}$1] projection and Figure \ref{figure:5}b shows a schematic simplified image at limited resolution. Since the twin boundary exhibits a mirror symmetry across a plane of tunnels (rather than atomic planes), the position of the mirror line in experimental HRTEM images may be used to determine whether the bright dots in an experimental image correspond to tunnels or atomic columns \cite{arne:spence}. The experimental HRTEM image in Figure \ref{figure:6} shows mirror symmetry across a line located between two rows of bright dots. Therefore the dots correspond to the atomic columns. The mirror symmetry across the two twin boundaries in the nanocrystal image in Figure \ref{figure:7} is present along a line with bright dots as indicated by arrows. Therefore the bright dots correspond to the tunnels in the silicon structure. In order to extract more information about the atomic structure of the nanocrystals a series of ten experimental HRTEM images were taken with a $\Delta$f=10 nm difference in focus. The focus of the recorded HRTEM images was determined experimentally. This was done by optical diffractograms by Fast Fourier Transformation (FFT), see Figure \ref{figure:8}. The method using optical diffractograms was developed by Thon in 1971 \cite{thon:gammel, Spence:gammel}. The experimental diffractograms from sample 70 showed three bright rings. The radial intensity of the ring pattern is approximately proportional to sin$^2 \chi$(u) where $\chi$(u) is the transfer function given by \cite{Spence:gammel}

\begin{equation}
\label{eq:1}
\chi (u) = (\pi \lambda \Delta f u^2) + (\frac{\pi}{2})(\lambda ^3 C_s u^4)
\end{equation}

$\lambda$ is the electron wave length, u is the distance to the maxima and minima in the optical diffractogram. C$_s$ is the spherical aberration coefficient, $\Delta$f is the defocus of the objective lens. The maxima or minima in the diffractograms correspond to sin$^2 \chi$(u)= 1 or 0. Equation \ref{eq:1} can be transformed to

\begin{equation}
\label{eq:2}
(n/{u^2})=(2 \lambda) \Delta f + (C_s \lambda ^3)u^2
\end{equation}

where n odd for maxima and n even for minima. A plot of y=n/u$^2$ as function of x=u$^2$ gives a line that intersect the y-axis at b=2$\lambda \Delta$f and with a slope a=C$_s \lambda^3$. For the experimental HRTEM images recorded in this work the electron wave length $\lambda$ was 0.00224nm as calculated for U = 300 kV and the instrumental value of the spherical aberration coefficient C$_s$= 0.6 mm was used. This gave a slope of a= 0.0067nm$^4$. The intersection with the y-axis b=2$\lambda \Delta$f was used to determine the defocus.

\begin{figure}
  \begin{center}
    \includegraphics[width=0.4\textwidth]{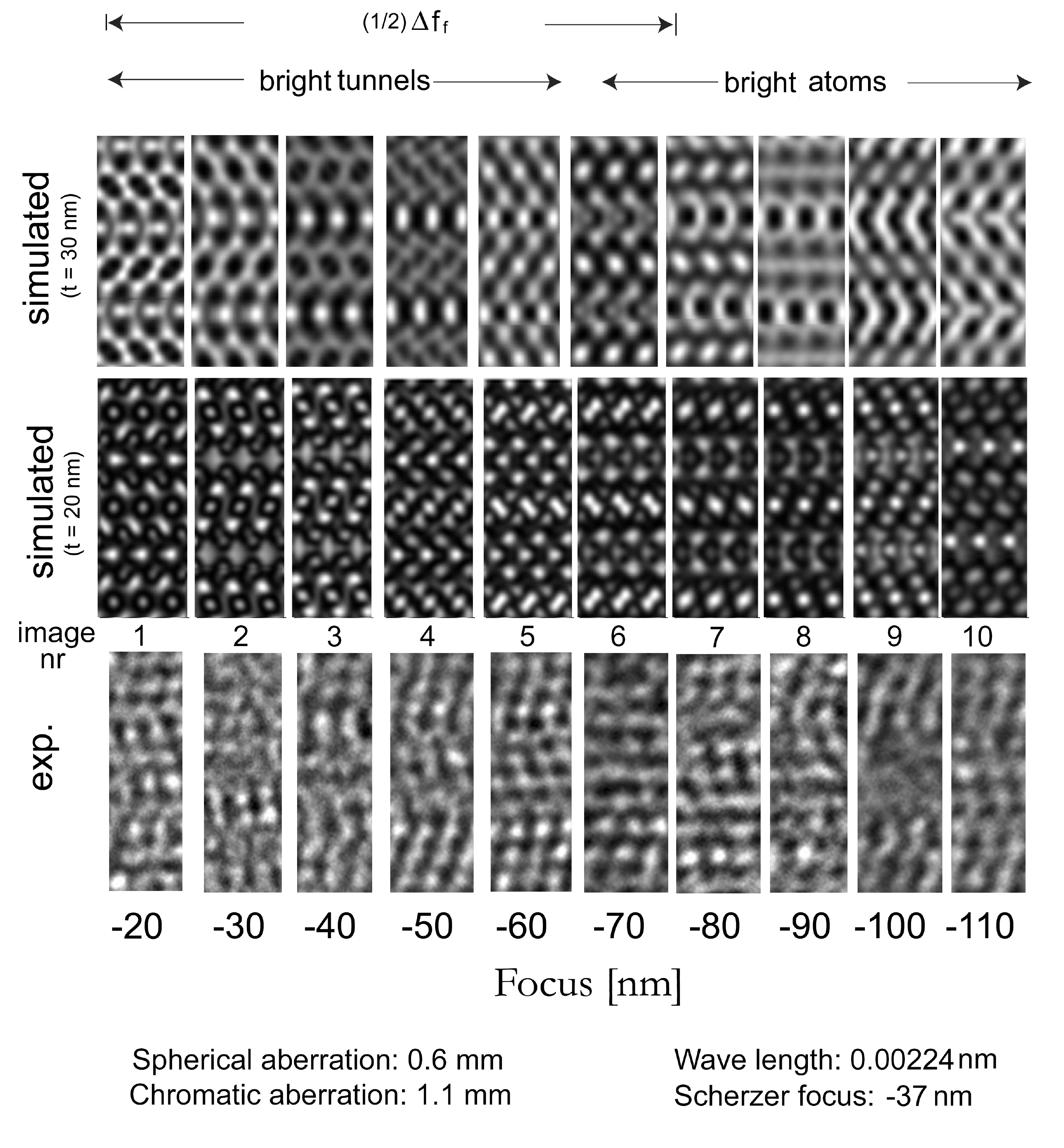}
    \caption{Experimental through focus series of the nanocrystal with a double twin structure as in Figure \ref{figure:7}, together with two series of simulated HRTEM images at different thickness and focus.}
    \label{figure:9}
  \end{center}
\end{figure}

The function $n/{u^2}$ versus u$^2$ as given by Equation \ref{eq:2} was plotted with n=-1 to n=-10. The curves are presented in Figure \ref{figure:8}. The square of the reciprocal distances (u) from the central beam to a maxima and minima of sin$^2 \chi$(u) in the optical diffractogram are plotted along the x-axis. The intersections between these points for a particular HRTEM image and the theoretical curves (with different n values), should satisfy the linear equation \ref{eq:2}. The two straight lines shown in Figure \ref{figure:8} were calculated with regard to HRTEM image 6 and 10 in Figure \ref{figure:9}. The defocus ($\Delta$f) was then determined by the intersection of the straight line with the y-axis b=2$\lambda \Delta$f. This procedure was carried out for each HRTEM image in the focus series, and was found to cover a variation in $\Delta$f from -20nm to -110nm. Unfortunately the measurements of the maxima and minima in the diffractograms have a relatively large standard deviation and this results in an accuracy of typically $\pm$10nm in $\Delta$f.

The experimental focus series were compared to simulated HRTEM images for different sample thicknesses with focus variations from $\Delta$f= -20nm to $\Delta$f= -110nm. The simulated through focus series were performed using the MacTempas computer program. The best agreement between the experimental and simulated images was found for thicknesses to be between 20 and 30 nm, see Figure \ref{figure:9}. This figure presents two series of simulated HRTEM images as a function of focus, calculated for sample thickness of 20 and 30 nm. These images were then used to identify the relationship between the image contrast and atomic structure near a twin boundary. From these images it is clear that the bright dots in the experimental HRTEM images with a focus $\Delta$f=-20nm to $\Delta$f=-60nm correspond to the tunnels. For the images with other focuses (-70 to -110) the bright dots correspond to the atomic columns as can be deduced from the position of the mirror line in the image.

The Fourier image period was calculated using equation \ref{eq:deltaf}, for various reflections ($u_{hkl}$) inside the objective aperture \cite{arne:spence}.

\begin{equation}
\label{eq:deltaf}
\Delta f_f = \frac{2}{\lambda u^2_{hkl}}
\end{equation}

The Fourier image period is the period at which the first and last HRTEM images in a series are identical. The point resolution of the microscope was calculated to be $X_{min}$=0.66$C_s^{1/4} \lambda^{3/4}$=0.174 nm. Lattice distances in Si larger than the point resolution of the microscope are $d_{111}$= 0.314nm, $d_{200}$= 0.272nm and $d_{220}$= 0.192nm. Therefore the objective aperture used for the experimental images was large enough to include these 3 types of reflections. The Fourier image periods corresponding to these reflections are $\Delta f_f(u_{111})$= 97.7nm, $\Delta f_f(u_{200})$= 73.8nm and $\Delta f_f(u_{220})$= 36.9nm respectively. By comparing these three periods with the HRTEM images in Figure \ref{figure:9}, the periods 36.9nm and 73.8nm do not satisfy the conditions of the Fourier image period. The first and last calculated HRTEM images in the two periods are not identical. This shows that the Fourier image period of the experimental image in Figure \ref{figure:9} is 97.7nm.

\subsection{Nanocrystal defects: Stacking faults and twinning}

\begin{figure}
  \begin{center}
    \includegraphics[width=0.4\textwidth]{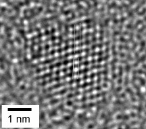}
    \caption{HRTEM image of a small (3nm) nanocrystal in sample 50.}
    \label{figure:10}
  \end{center}
\end{figure}

\noindent Nanocrystals smaller than 3 nm in diameter show large lattice strains on planes, see Figure \ref{figure:10}. The distorted planes might be due to surface effects when the Si nancrystal are embedded in the SiO$_2$ matrix. During nanocrystal growth, this strain might lead to both twinning and stacking faults. 

In the crystalline nanoclusters twinning and stacking faults occur, even in clusters down to 4 nm in size. In large elongated nanocrystals no defects were observed. Figure \ref{figure:6} presents a 7.5 nm nanocrystal with a single twin structure, as seen in the [0$\overline{1}$1] projection. The twin plane is indicated by an arrow labeled TB. The angle between the atomic planes on opposite sides of the twin boundary is 70.5$^o$, which agrees well with the value 70.53$^o$ between the $\{$111$\}$-type planes in cubic materials.  The projected shape of the nanocrystal is hexagonal. This is in agreement with the report by Wang et al. \cite{wang:twin}, who also found a hexagonal shape of nanocrystals containing a single twin boundary. The nanocrystal boundary consists of $\{$111$\}$ and $\{$110$\}$ crystal facets. The parallel white lines illustrate a deviation in the atomic stacking characteristic for stacking faults. 

Figure \ref{figure:7} presents a 4 nm nanocrystal with a double twin structure and two stacking faults. The twin planes are $\{$111$\}$ planes and the twin boundaries are indicated by arrows labeled TB1 and TB2. Facets are visible along the $\{$111$\}$ planes. The two twin boundaries are parallel and only separated by 4 atomic tunnels. The nanocrystal stacking faults are located in the twin plane and is passing through the whole nanocrystal. Both twins and stacking faults frequently occur in the equivalent plane, the $\{$111$\}$ plane, in diamond type structures \cite{amelinckx:si}. The atomic sequence of the stacking fault and twin (ABCBACBA) is shown on Figure \ref{figure:7}. The stacking fault is extrinsic, and can be formed by inserting an extra atomic plane B in the perfect crystal (ABCABC to ABC\textbf{B}A). Extrinsic stacking faults can be described by use of shears in successive planes in a perfect crystal. Wang et al.\cite{wang:twin} proposed that when the stacking faults and twinning coexist, the projected shape becomes more irregular. The projected shape of the nanocrystals in the present work was not irregular. This may lead to the conclusion that when stacking fault is located in the twinning planes, the projected shape is the same as in a stacking fault free crystal.

Contrary to the samples studied by Wang et al.\cite{wang:twin} made by ion implantation, twinning and stacking faults were found in nanocrystals smaller than 5 nm. Perrey et al.\cite{perry:defect} studied silicon nanocrystals embedded in amorphous hydrogenated silicon. They also found twinning in nanocrystals smaller than 5 nm in diameter. The sample preparation method and matrix type around the silicon nanocrystals seem to affect the defect concentration in the nanocrystals. Increased SiO$_x$ in the matrix may imply higher surface stresses to the nanocrystals and may induce twinning and stacking faults.

\section{Conclusion}

\noindent Nanocrystals of silicon in a thin SiO$_2$ matrix upon a Si substrate were studied with HRTEM, EFTEM and SIMS. The nucleation and their structure in silicon were examined in detail. The studies show that nanocrystals precipitate at a silicon fraction of 50 area \% and above. Below this percentage amorphous silicon nanoclusters were found. Both crystalline and amorphous nanoclusters precipitate in a 4 nm band, 5 nm from the silicon substrate and 4 nm from the specimen surface. EELS, EDS and SIMS show a higher silicon concentration in the RTO layer and more oxygen in the top layer. These results may indicate that oxygen diffuses into the oxide close to the surface, whereas silicon diffuses from the SiO$_2$-Si layer and into the RTO during heat treatment. The defects in the nanocrystals were studied in detail, and found to be twinned clusters with extrinsic stacking faults. The smallest nanocrystal with defects is 4 nm in diameter. The small amorphous nanoclusters have a spherical shape while the crystalline nanocrystals have a more hexagonal shape with facets. After further heat treatment, higher temperature or silicon concentration, the nanocrystals have an elongated growth in the $<$111$>$ direction. Detailed HRTEM image simulations were carried out to get more information about the nanocrystals and their defects. All stacking faults found in the nanocrystals are extrinsic, going through the whole nanocrystal and is located in the twin boundary. No dislocations were observed.

\section{Acknowledgement}

\noindent Financial support by FUNMAT@UiO, the University of Oslo, Kristine Bonnevie's and SCANDEM's travelling scholarship is gratefully acknowledged.

\end{document}